\documentclass[twocolumn,superscriptaddress,aps,pre,bibtex]{revtex4-1}
\usepackage[T1]{fontenc}
\usepackage[latin9]{inputenc}

\usepackage[table,svgnames]{xcolor}
\usepackage{array}
\usepackage{units}
\usepackage{amsmath}
\usepackage{amssymb}
\usepackage{graphicx}
\usepackage[normalem]{ulem}
\usepackage{filecontents}

\begin{document}

\title{Self-organized bistability and its possible relevance for brain
  dynamics}

\author{Victor Buend{\' i}a} \affiliation{Departamento de
  Electromagnetismo y F{\'i}sica de la Materia e Instituto Carlos I de
  F{\'i}sica Te{\'o}rica y Computacional. Universidad de Granada,
  E-18071 Granada, Spain} \affiliation{Dipartimento di Scienze
  Matematiche, Fisiche e Informatiche, Universit\`a di Parma, via
  G.P. Usberti, 7/A - 43124, Parma, Italy} \affiliation{INFN, Gruppo
  Collegato di Parma, via G.P. Usberti, 7/A - 43124, Parma, Italy}
\author{Serena di Santo} \affiliation{Morton B. Zuckerman Mind Brain Behavior Institute
Columbia University, NY, USA} \author{Pablo Villegas} \affiliation{Istituto dei
  Sistemi Complessi, CNR, via dei Taurini 19, 00185 Rome, Italy}
\author{Raffaella Burioni} \affiliation{Dipartimento di Scienze
  Matematiche, Fisiche e Informatiche, Universit\`a di Parma, via
  G.P. Usberti, 7/A - 43124, Parma, Italy} \affiliation{INFN, Gruppo
  Collegato di Parma, via G.P. Usberti, 7/A - 43124, Parma, Italy}
\author{Miguel A. Mu\~noz} \affiliation{Departamento de
  Electromagnetismo y F{\'i}sica de la Materia e Instituto Carlos I de
  F{\'i}sica Te{\'o}rica y Computacional. Universidad de Granada,
  E-18071 Granada, Spain} \affiliation{Dipartimento di Scienze
  Matematiche, Fisiche e Informatiche, Universit\`a di Parma, via
  G.P. Usberti, 7/A - 43124, Parma, Italy}

\begin{abstract}
  Self-organized bistability (SOB) is the counterpart of
  ``self-organized criticality'' (SOC), for systems tuning themselves
  to the edge of bistability of a discontinuous phase transition,
  rather than to the critical point of a continuous one.  The
  equations defining the mathematical theory of SOB turn out to bear
  strong resemblance to a (Landau-Ginzburg) theory recently proposed
  to analyze the dynamics of the cerebral cortex. This theory
  describes the neuronal activity of coupled mesoscopic patches of
  cortex, homeostatically regulated by short-term synaptic plasticity.
  The theory for cortex dynamics entails, however, some significant
  differences with respect to SOB, including the lack of a (bulk)
  conservation law, the absence of a perfect separation of timescales
  and, the fact that in the former, but not in the second, there is a
  parameter that controls the overall system state (in blatant
  contrast with the very idea of self-organization).  Here, we
  scrutinize --by employing a combination of analytical and
  computational tools-- the analogies and differences between both
  theories and explore whether in some limit SOB can play an important
  role to explain the emergence of scale-invariant neuronal avalanches
  observed empirically in the cortex. We conclude that, actually, in
  the limit of infinitely slow synaptic-dynamics, the two theories
  become identical, but the timescales required for the
  self-organization mechanism to be effective do not seem to be
  biologically plausible.  We discuss the key differences between
  self-organization mechanisms with/without conservation and
  with/without infinitely separated timescales. In particular, we
  introduce the concept of ``self-organized collective oscillations''
  and scrutinize the implications of our findings in neuroscience,
  shedding new light into the problems of scale invariance and
  oscillations in cortical dynamics.
  \end{abstract}
\maketitle

\section{Introduction}
The theory of self-organized criticality (SOC) explains how systems
can become self-organized to the edge of a continuous phase
transition, i.e. to the vicinity of a critical point without the
apparent need of parameter fine tuning
\cite{Bak,BTW,Pruessner,Jensen,BJP}.  This theory (or mechanism) is
often invoked to explain the emergence of scale-free distributions in
the sizes and durations of outbursts of activity --often called
``avalanches''-- interspersed by periods of quiescence in diverse
settings such as earthquakes, vortices in superconductors
\cite{Bak,Pruessner,BJP}, and cortical brain activity
\cite{Chialvo2004,Levina2007,Millman,RMP}, to name but a few.

The basic mechanism for self-organization to the edge of a phase
transition --as exemplified by its most paradigmatic representatives,
sandpile models \cite{BTW,Manna,Oslo}-- relies on two essential and
intertwined features: The first one is the presence of two infinitely
separated timescales: a fast dynamics characterizes the system
intrinsic activity, while a slow dynamics modulates the control
parameter. This latter acts differentially on each phase --at
  each side of the transition-- thus creating a feedback loop that
  self-organizes the system to the edge of the transition
  \cite{Sontag,GG2,BJP,JABO1}).  The second important feature is that
the bulk dynamics is conserved, i.e. dissipation occurs only at the
system boundaries: in the presence of bulk dissipation there
  would be characteristic timescales --both in space and in time--
  incompatible with the idea of scale invariance.

Nevertheless, self-organized models lacking conservation --such as
forest-fire and earthquake models \cite{FFM,Olami}-- have also a long
tradition in studies of self-organized criticality. The main
  difference between conserved and non-conserved dynamics in SOC
  models is that while the first one drives the dynamics exactly to
the critical point with concomitant scale invariance, non-conserved
dynamics leads to a wandering around the critical point (i.e. with
excursions to both sides of the critical point but not sitting exactly
on it). This mechanism has been termed ``self-organized quasi
criticality'' (SOqC) and leads to approximate scale invariance for a
few decades, which may suffice to describe what empirically observed
in real systems, without the need to invoke perfect criticality as in
SOC (we refer the reader to \citet{JABO1} for a thorough and
pedagogical discussion of these issues).

The paradigm of self-organized criticality has been very influential
in the context of neuroscience. It has been profusely employed to
rationalize the empirical observation of scale-free avalanches,
i.e. outbursts of neuronal activity whose sizes and durations are
distributed as power laws \cite{BeggsPlenz}, as robustly found across
species, scales, and experimental techniques \cite{Schuster,Cocchi}.
Mathematical models inspired in SOC have been proposed to account for
such scale-invariant avalanches
\cite{Levina2007,Millman,Lucilla2006,JABO2}.  Actually, the idea that
the cerebral cortex --as well as other biological systems-- might
extract important functional benefits from operating at criticality,
has attracted a lot of interest and excitement as well as some
controversy (see e.g. \cite{Schuster,Chialvo2010,KC,Cocchi,RMP}); SOC,
as well as SOqC, played a key role in the development of this
conjecture \cite{JABO2,Kinouchi}.

Recently, a mechanism similar in spirit to SOC --relying also on an
infinite separation of timescales and on a conserved dynamic for the
control parameter-- has been discovered to be able to self-organize
systems exhibiting a discontinuous phase transition to the edge of
such a transition.  As a matter of fact, self-organized bistability
(SOB) --as the theory has been named-- has been proposed as a new and
very general paradigm for the self-organization to points of
bistability or phase coexistence between two alternative phases \cite{SOB}.

Systems under SOB conditions turn out to be also characterized by
scale-free avalanches of activity. More specifically, localized small
perturbations into the quiescent state propagate in the form of
avalanches with sizes $S$ and durations $T$ distributed as power laws:
$P(S)\sim S^{-\tau}$ and $P(T)\sim T^{-\alpha}$, and the averaged
avalanche size scales as $\langle S\rangle\sim T^{\gamma}$, obeying
the scaling relation $\gamma=(\alpha-1)/(\tau-1)$
\cite{Sethna,Avalanches}. However, at odds with SOC, avalanches in SOB
systems are distributed in a bimodal fashion, i.e. for any finite
system size their sizes and times distributions consist of a power-law
complemented with a ``bump'', corresponding to anomalously large
system-spanning events \cite{SOB,Kinouchi}. Furthermore, the exponents
$\tau, \alpha$ and $\gamma$ differ from their SOC counterparts and
coincide with those of the usual mean-field branching process
exponents even in low dimensional systems
\cite{SOB,Harris,LiggettInteracting,WG}.
  
Given that living systems could have evolved to exploit the
  complementary benefits of two alternative coexisting phases
  \cite{RMP}, this new paradigm for the self-organization to the edge
  of a phase transition has been argued to be of potential relevance
  in biological problems, much as SOC is.  In particular, in the
original paper in which SOB was introduced, it was also speculated
that SOB might be relevant to explain the emergence of scale-free
avalanches in the neuronal activity of the cerebral cortex \cite{SOB}.
 
During deep sleep or under anesthesia, the state of cortical activity
is well-known to exhibit a form of bistability, with an alternation
between two possible states of high and low levels of neural activity
--called ``\emph{up} and \emph{down}'' states--, respectively
  \cite{Tsodyks, Curto2009, Compte2003, Bazhenov2002}. This
underlying bistability, together with the empirical observation of
scale-free avalanches --appearing sometimes in concomitance with
anomalously large outbursts-- in the awake resting brain, made the
authors of Ref~\cite{SOB} suggest (as a possibility to explore) that
there could be some type of self-organization to the edge of
bistability, rather than the usually postulated self-organization to
criticality \cite{Levina2007,Millman,Lucilla2006,JABO2} or any other
alternative scenario.

In a parallel endeavor, our research group has recently proposed a
physiologically-motivated mesoscopic (Landau-Ginzburg) theory,
specifically designed to shed light on the large-scale dynamical
features of cortical activity \cite{PNAS-Synchro}. The outcome of such
an approach is that the relevant phase transition for cortical
dynamics is a synchronization phase transition (which occurs in
concomitance with scale-invariant avalanches) with no
self-organization to such a transition: parameters need to be fine
tuned to observe it.  Moreover, the resulting synchronous-asynchronous
phase transition is of a different nature of the quiescent-active
phase transition assumed to describe avalanches in cortical dynamics,
which has motivated a change of perspective in the field with
important consequences.

The most remarkable fact for our purposes here is that the
Landau-Ginzburg theory for cortical dynamics bears profound
similarities with that of SOB. Thus, our main goal here is to
scrutinize the analogies and differences between the above
Landau-Ginzburg theory and the theory of SOB.  In particular, we pose
the following question: can self-organization to the very edge of a
discontinuous transition with scale-free avalanches be possibly
observed in the (Landau-Ginzburg) model? Can the SOB theory be
modified to reproduce the phenomenology of the Landau-Ginzburg
equation? Answering these questions will pave the way to a deeper
understanding of self-organization mechanisms and their relevance in
neuroscience.

\section{The theory of self-organized bistability}

Let us overview here the main aspects of the theory of SOB (for the
sake of clarity and self-containment, we present in Appendix A a
simple one-site or mean-field formulation that might be helpful to
gain insight for readers not familiar with SOB or SOC theories; for a
complete description of SOB and its mathematical formulation we refer
to \citet{SOB}).

Given a spatially-extended system, the theory of SOB can be written in
terms of two equations, one for an activity field $\rho(\vec{x},t)$ (which
in sandpiles represents sites over the instability threshold)
and one for a ``background'' or ``energy'' field (which in
  sandpiles represents the total amount of grains at a given site)
\footnote{Note that either name is an abstraction, e.g. in sandpiles
  this field represents the local amount of sandgrains.}.

More precisely, the set of Langevin equations proposed to describe the
evolution of the activity $\rho({\vec{x}},t)$ and ``energy'' field
$E({\vec{x}},t)$ in the SOB theory \cite{SOB}, read
\begin{equation}
  \begin{array}{lll}
    \dot{\rho}(\vec{x},t) = [E (\vec{x},t)-a]\rho+b\rho^{2} - \rho^3 
    +D\nabla^2\rho+ h +\eta({\vec{x}},t)\\ 
    \dot{E}({\vec{x}},t) = \nabla^2\rho (\vec{x},t)-\varepsilon
    \rho (\vec{x},t) +h,
\end{array}
\label{SOB}
\end{equation}
(note that some dependencies on
$(\vec{x},t)$ have been omitted in Eq.(\ref{SOB}) for simplicity),
$a, b>0$ are constants, $\nabla^{2} \rho$ stands for a diffusive
coupling with diffusion constant $D$, $\eta (\vec{x},t)$ is a
zero-mean multiplicative Gaussian noise with
$\langle \eta({\vec{x}},t) \eta({\vec{x'}},t') \rangle =
\rho(\vec{x},t)\delta(\vec{x}-\vec{x'}) \delta(t-t')$ describing
particle-number (``demographic'') fluctuations. 

This is the simplest set of equations extending the theory of
  self-organized criticality \cite{FES-PRL,FES-PRE,BJP,JABO1} to the
  case in which there is a discontinuous phase transition
  \cite{Eluding}.  More in particular, the rationale behind these
equations is as follows. The first one for the field activity exhibits
a discontinuous phase transition for the overall level of activity as
the control parameter (e.g. value of $E$) is changed. The second one
is responsible for creating a feedback loop between the control and
the order parameters \cite{Sontag}.  When there is no activity, the
local energy field grows with the constant driving $+h$ (shifting the
system towards the active phase) while, in the presence of activity,
dissipation at the boundaries dominates (shifting the system towards
the absorbing phase).  Importantly, $h$ is an arbitrarily small
driving rate, which, in analogy with the driving in sandpile models
--where just one sandgrain is added at the time \cite{Bak}-- is
assumed to scale with the inverse of the system size $h\propto1/L^2$.
Equivalently $\varepsilon$ is the rate of the activity-dependent
energy dissipation, which in sandpile models is a boundary effect,
scaling as $\varepsilon\propto1/L$; thus, the ratio
$h/\varepsilon \rightarrow 0$ in the large system-size limit. In other
words, the SOB limit is obtained when driving and dissipation go to
zero, i.e. the energy equation converges to be conserved and the
timescales of both equations become infinitely separated (in the
jargon of non-linear systems this corresponds to a ``slow-fast''
dynamics \cite{Desroches}).

As mentioned in the introduction, the outcome of this set of equations
is that the system self-organizes to the point of bistability between
the quiescent and the active phase with scale-free avalanches
\cite{SOB}.  In particular, the associated exponents $\tau$ and
$\alpha$ and $\gamma$ take well-known values, i.e. those of an
unbiased branching process \cite{Harris,SOBP,Branching} (actually, to
be more precise, they coincide with those of compact directed
percolation or voter model universality class in all dimensions
\cite{Avalanches,Marro,Henkel}, and with those of the branching
process above the upper critical dimension $d=2$ \cite{SOB}). Also, as
said above, such scale-free distributed avalanches coexist (for any
finite system size) with anomalously large events or waves, in which
activity spreads ballistically, weeping the whole system \cite{SOB}.

\section{The Landau-Ginzburg theory of cortex dynamics}

The set of Eqs.(\ref{SOB}) can be directly compared with the
physiologically-inspired Landau-Ginzburg model that has been recently
proposed to shed light on the large-scale features of brain activity,
relying on synaptic plasticity as a chief regulatory mechanism
\cite{PNAS-Synchro}.  In this modelling approach, neural activity is
described at a coarse-grained level in the spirit of the approach of
\emph{Wilson and Cowan} to large-scale neural dynamics
\cite{WC,WC-review}. On the other hand, short-term synaptic plasticity
is implemented as a main regulatory mechanism --using the celebrated
\emph{Tsodyks-Markram} model \cite{Tsodyks} (see also the very similar
approach by Abbott \emph{et al.} \cite{Abbott1997})-- in line with previously
proposed models of self-organization in brain dynamics
\cite{Levina2007,Levina2009,Millman,Lucilla}. In particular, the level
of neural activity at each coarse-grained or mesoscopic region of the
cortical tissue (representing a local subpopulation of neurons) is
encoded in an activity variable $\rho(\vec x, t)$, while the local
amount of available synaptic resources is called $R(\vec x, t)$.  The
following set of equations defines the dynamics of these two variables
\cite{PNAS-Synchro}:
\begin{equation}
\begin{cases}
  \dot{\rho}(\vec x, t)=(R(\vec x, t)-a)\rho+b\rho^2 -\rho^{3}+I
  + D\nabla^{2}\rho+ \eta(\vec x, t)\\
  \dot{R}(\vec x, t)=\frac{1}{\tau_R}(\xi-R)- \frac{1}{\tau_D} R\rho
\end{cases}
\label{PNAS}
\end{equation}
(where, again, some dependences on $(\vec x, t)$ have been omitted for
the sake of simplicity).  In the equation for the activity field
--much as above-- $a, b>0$ are constant parameters, $I$ stands for a
small external incoming input, $\nabla^{2} \rho$ stands for the
diffusive coupling between local regions with diffusion constant $D$,
and $\eta(\vec x, t)$ is a zero-mean Gaussian white noise.  The
bilinear coupling between $R$ and $\rho$ in the first equation
reflects the fact that the larger the amount of synaptic resources,
the larger the rate at which further activity is generated.  On the
other hand, in the second the equation, $\tau_R$ and $\tau_D$ are the
characteristic scales for the processes of recovery and depletion of
synaptic resources, respectively, $\xi$ is the maximal level of
available synaptic resources that can possibly be reached.

Let us emphasize that --importantly for what follows-- this
  theory is not self-organized, i.e. the resulting phenomenology
depends on the value of the baseline level of synaptic resources,
$\xi$:

\begin{itemize}
\item For small values of $\xi$ the system sets into the quiescent (or ``down'')
  state, with vanishing activity.
\item For large values of $\xi$ the system sets into an active (or
  ``up'') state with a sustained (high) level of activity, which is
  sometimes called ``asynchronous irregular'' phase.
\item In between the above two regimes there is an intermediate one
  called ``synchronous irregular'' in which there are intermittent
  events which are synchronized; these correspond to fast waves
  traveling through the system, and generating co-activation of many
  units within a relatively small time window.
\item Right at the critical point separating the synchronous from the
  asynchronous phase avalanches can be measured by using a protocol
  --relying on the definition of a discrete time binning-- identical
  to that used by experimentalists to detect neuronal avalanches
  \cite{BeggsPlenz,RMP}.
\end{itemize}

At the light of these facts, it was proposed that the actual
scale-free avalanches of neuronal activity observed in the cortex stem
from the dynamics operating in a regime close to the
synchronous-asynchronous phase transition, rather than to the critical
point of a standard quiescent-to-active phase transition as had been
assumed before \cite{Schuster}. However, the mechanism by which the
cortex seemingly selects to operate near such a critical point remains
undetermined \cite{PNAS-Synchro}.

\section{Analogies and differences between both theories}
Remarkably, the sets of equations Eq.(\ref{SOB}) and Eq.(\ref{PNAS})
exhibit profound analogies (note that the notation in both of them has
been fixed to make the similitudes self-evident).  As a matter of
fact, the first equation, which describes in the simplest possible
(minimal) way a discontinuous/first-order phase transition into an
absorbing state, is identical for both theories. On the other hand,
formal differences appear in the second equation for the ``energy''
field in SOB theory, Eq.(\ref{SOB}), which in Eq.(\ref{PNAS}) is
replaced by a field describing the dynamics of synaptic resources.

A first important difference between both sets of equations is that in
SOB the energy field is conserved in the bulk; only driving events and
boundary dissipation make the total integral value fluctuate in time,
but as specified earlier, they are both vanishingly small.  On the
other hand, the above equation for $\dot{R}(\vec x, t)$ is not
conserved in the bulk, as it includes a (positive) term for the
charging/recovery of resources to their baseline level $\xi$, as well
as a (negative) term for the activity-dependent consumption of
resources.

The second (related) difference is that in Eq.(\ref{PNAS}) there is no
perfect separation of timescales. Actually, the synaptic time
constants $\tau_R$ and $\tau_D$ are biologically motivated parameters,
e.g. $\tau_R=10^3$ and $\tau_D=10^2$ time units \cite{PNAS-Synchro},
that cannot be assumed to scale with sytem size (as would be required
in SOB) \cite{JABO2}.

In third place, in Eq.(\ref{PNAS}) there is a baseline level of
synaptic resources $\xi$ which, as said above, plays the role of an
overall control parameter: by tuning it one can shift the system
between different phases from a quiescent regime with no activity, to
one with oscillations, waves etc. (see \cite{PNAS-Synchro}); such a
tuning parameter has no counterpart in SOB, Eq.(\ref{SOB}). Indeed,
the very existence of a control parameter allowing to shift between
dynamical regimes is at odds with the very idea of self-organization.

Thus our aim in what follows is to explicitly explore whether
Eq.(\ref{PNAS}) can lead to the same behavior of SOB --i.e. to true
self-organization-- in some limit and, also, the other way around: to
see what modifications of SOB lead to the Landau-Ginzburg
phenomenology.

\begin{figure}[h]
\centering
\includegraphics[width=1\columnwidth]{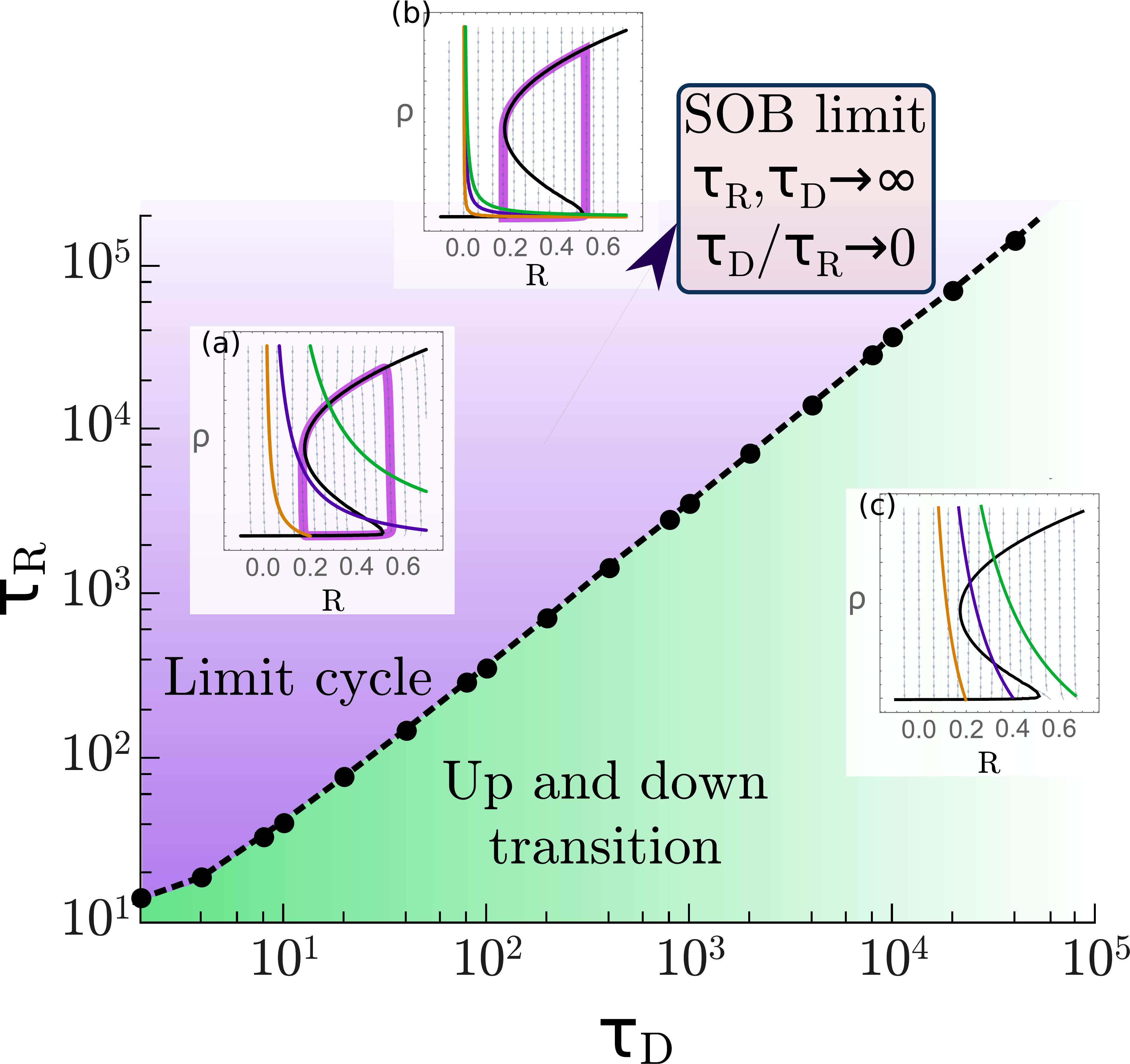}
\caption{{\bf Phase portraits and nullclines} for the deterministic
  dynamics at an individual site of the Landau-Ginzburg theory,
  Eq. (\ref{PNAS}).  Insets: the nullcline for $\dot{\rho}=0$ is
  colored in black, while nullclines for $\dot{R}=0$ --for three
  different values of the baseline level of synaptic resources $\xi$--
  are plotted in orange, purple and green color, respectively. The
  small grey arrows represent the vector field for $(\dot\rho,\dot R)$
  and the light purple curve represents a limit-cycle
  trajectory. \textbf{Inset (a)} For $\tau_D/\tau_R=0.1$ and
  $\tau_R=10^3$ the system may display a down state (orange nullcline,
  $\xi=0.2$), a limit cycle (purple nullcline, $\xi=1.5$) or an up
  state (green nullcline, $\xi=3.5$). \textbf{Inset (b)} For
  $\tau_D/\tau_R=0.001$ and $\tau_R=10^7$ the system approaches the
  SOB behavior and the dependence on the control parameter $\xi$
  becomes very weak, meaning that for a wide range of values of the
  control parameter above the down state ($\xi=1,5,10$ for orange,
  purple and green nullclines respectively) the system sets in the
  oscillatory phase. \textbf{Inset (c)} When the separation of
  timescales is very low, the system only displays bistability between
  up and down states with no oscillations whatsoever, no matter the
  value of $\xi$. \textbf{Main} The global behavior of the system
  depending on the timescale separation is coded in the main figure
  with different background colors: purple when the limit cycle
  appears, and green when only bistability is possible.  Parameters
  are set to $I=10^{-3}, a=0.6, b=1.3$, in both cases.}
\label{fig:mf}
\end{figure}

\section{Results}
Let us start with a single unit (or mean-field) analysis of the
local/individual components of Eq. (\ref{PNAS}) in the stationary
state, for which one needs to detect where the two nullclines of the
associated deterministic equations (for the one-site dynamics)
intersect (see Appendix B for more details as well as Appendix C for a
related model). The fixed points $ (\rho^*,R^*)$ of the dynamical
system depend on the maximal level of synaptic resources, $\xi$. For
small values of $\xi$, the deterministic system settles in a ``down''
state with $\rho^*=0$ and $R^* = \xi$. In contrast, an ``up'' state
with sustained activity, $\rho^*>0$, with a depleted level of
resources, $R^* <\xi$, emerges for sufficiently large values of $\xi$.

Separating these two limiting states, there is a range of values of
$\xi$ where a stable limit cycle emerges if the timescale separation
is large enough (see leftmost inset of Fig.\ref{fig:mf}a and
purple-shaded region in Figure \ref{fig:mf}).  As discussed in detail in
  Appendix B, the limit cycle is created via a homoclinic bifurcation
  at $\xi = a$ and destroyed via a Hopf bifurcation at a value of
  $\xi$ that depends on the timescales. In particular, for a
biologically plausible separation of the characteristic timescales
$\tau_R$ and $\tau_D$ (i.e.  $\tau_D/\tau_R=0.1$ and $\tau_R=10^3$ as
chosen in \cite{PNAS-Synchro}) the system exhibits an intermediate
regime with oscillations as illustrated in Figures \ref{fig:mf}a and Appendix B. These
oscillations at the individual sites are at the origin of the
propagating waves and emerging synchronous behavior in the spatially
extended system \cite{PNAS-Synchro}.

\begin{figure*}
\centering
\includegraphics[width=1.5\columnwidth, height=5cm]{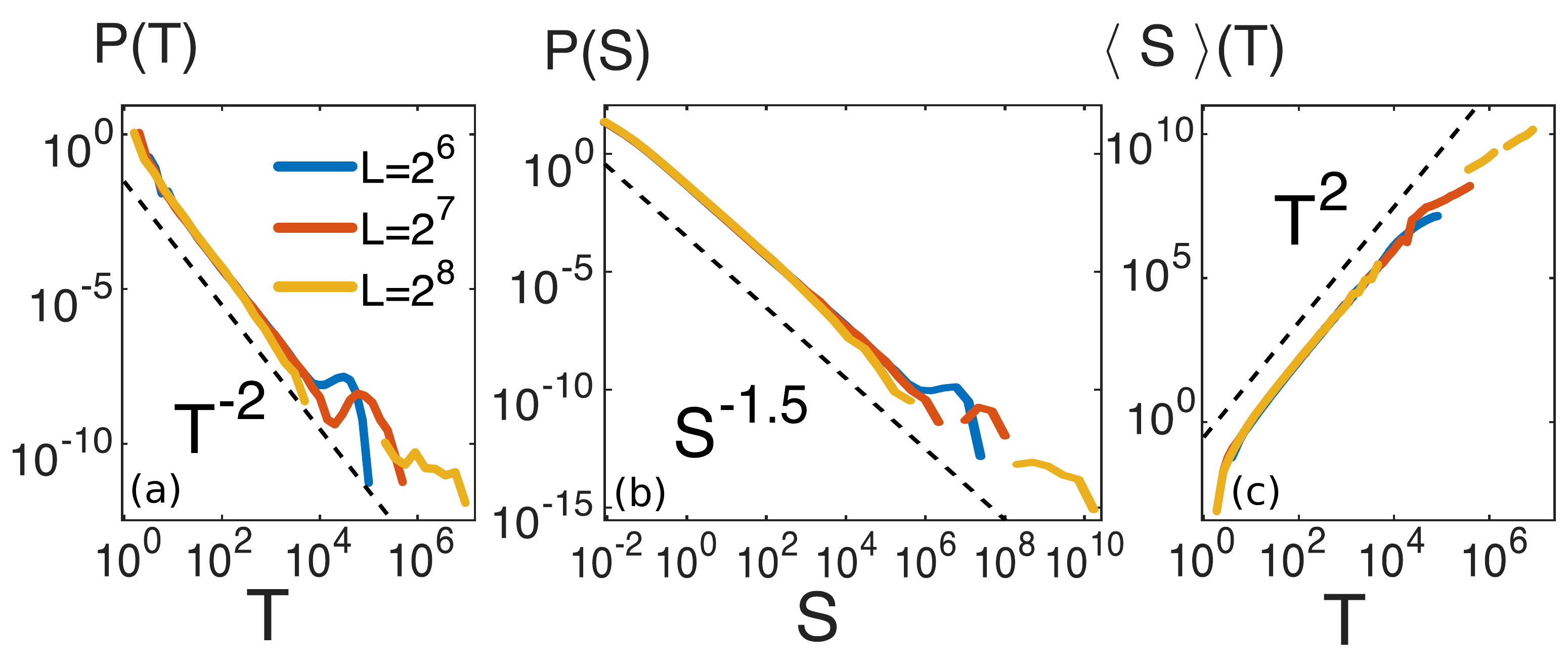}
\caption{{\bf Avalanches} in the limit of a large separation of
  timescales, for the spatially extended (two dimensional) version of
  the Landau-Ginzburg theory for cortical dynamics, Eq. (\ref{PNAS}).
  Probability distribution for avalanche durations $T$ (a),
  avalanche sizes $S$ (b) and average avalanche size as a
  function of duration (c) in double logarithmic scale, for
  square-lattice systems of sizes: $N=2^{12}, 2^{14}$ and
  $2^{16}$. The dashed lines are plotted as a guide to the eye, and
  have slopes corresponding to the expectations for an unbiased
  branching process ($\alpha=-2, \tau=-3/2$ and $\gamma=2$,
  respectively). The ``bumps'' correspond to anomalously large events,
  i.e. synchronized spiking events. The cut-offs/bumps change with $N$
  obeying finite-size scaling as in the theory of self-organized
  bistability \cite{SOB}.  Parameters: $b=0.5$ $a=1$, $I=10^{-7}$,
  $\tau_D=10^4$, $\tau_R=10^6$, $D=1$.}
\label{fig:ava}
\end{figure*}

\begin{figure*}
\centering
\includegraphics[width=1.7\columnwidth, height=5cm]{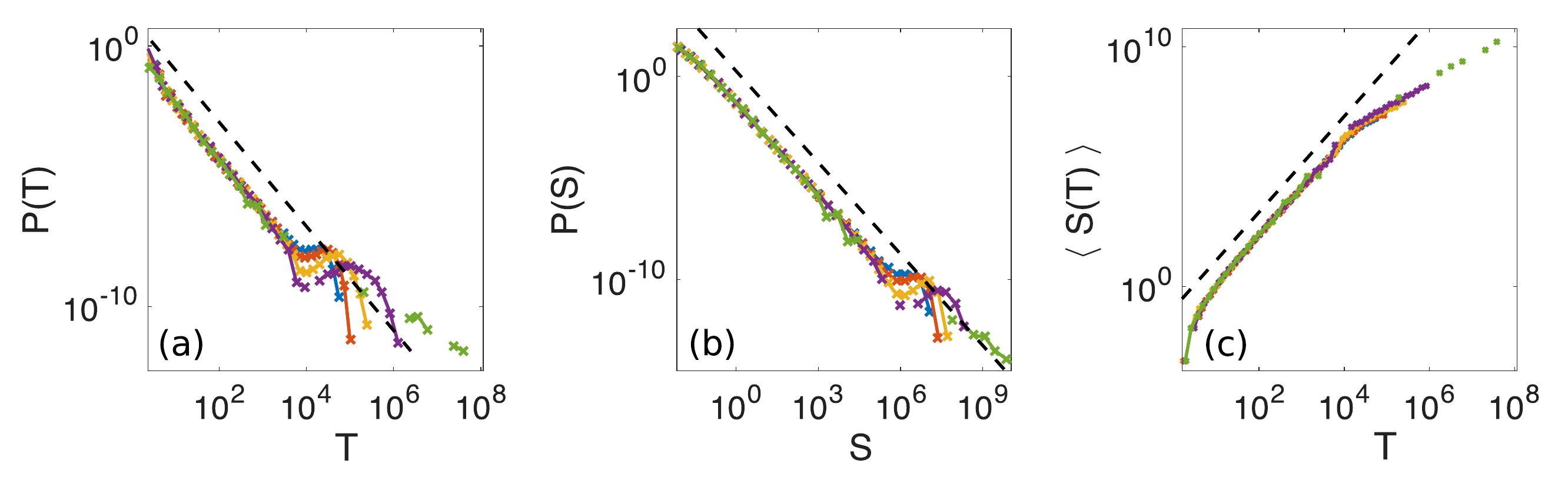}
\caption{{\bf Weak dependence on the control parameter} in the limit
  of large separation of timescales, for the spatially extended (two
  dimensional) version of  the Landau-Ginzburg theory for cortical dynamics, Eq. (\ref{PNAS}).
Different colors represent different values of $\xi$.  There
  is essentially no visible change in the probability distributions
  for avalanche durations $T$ (a), avalanche sizes $S$ (b), and
  average avalanche size as a function of duration (c), for
  different values of $\xi$, in particular,
  $\xi=1, 3, 5, 7, 10$. The dashed lines are plotted as a guide to the eye
  and their slopes correspond to unbiased branching process exponents.
  Parameters are fixed as in Fig.\ref{fig:ava} with $N=2^{14}$}
\label{fig:xis}
\end{figure*}
On the other hand, in the opposite limit in which the timescales are
both relatively fast and not very separated --a case that we do not
discuss here in further detail-- there is no intermediate regime of
oscillations and the system just shifts from a quiescent (down) to an
active (up) state as $\xi$ is increased (see the green-shaded region
of Fig.\ref{fig:mf} and Fig.\ref{fig:mf}c). This regime is likely to be useful to
  describe up-and-down transitions as observed in the cerebral
  cortex.


Given that the mechanism of the recovery of synaptic resources (with a
characteristic timescale $\tau_R$) plays the role of driving, while
the depletion of resources (with a characteristic scale $\tau_D$)
plays the role of dissipation in presence of neural activity, we
conjecture that the SOB limit of infinite separation of timescales
could be only possibly recovered taking formally the double limit
$1/\tau_R\rightarrow0$ and $1/\tau_D\rightarrow0$, with
$\tau_D/\tau_R\rightarrow0$. In particular, in what follows we
consider different sets of values for the timescales $\tau_R$ and
$\tau_D$ --regardless of their real meaning in neuroscience--
approaching a large separation of timescales (e.g.
$\tau_D/\tau_R=0.001$ rather than $\tau_D/\tau_R=0.1$ as above; see
Fig.\ref{fig:mf}b).

First of all, we observe that the larger the separation of timescales,
the weaker the dependence on the control parameter $\xi$ (see in
particular in Fig.\ref{fig:mf}b how the three different $\dot{R}=0$
nullclines --for three different values of $\xi$-- become very close
to each other). This allows us to reproduce one of the main features
of the self-organization mechanism: robustness on changes of the
tuning parameter emerges as the limit of infinite separation of
timescales is approached (see Appendix B for more details).

Second, when the separation of timescales tends to infinity, the slope
of the nullcline $\dot R=0$ converges to zero at the intersection
point.  Then, both nullclines intersect in a
tangential way, giving rise to a limit cycle, much as in the
mean-field SOB (see Appendix A).

To go beyond the single-site or mean-field analysis we now study
computationally a full (spatially explicit) system.  Under the
conditions of a large separation of timescales we have performed
computer simulations of the full set of Eqs.(\ref{PNAS}).  In
particular, we considered square lattices with sizes from $N=2^{12}$
to $N=2^{16}$ to analyze finite size effects and performed up to
$10^8$ runs for different parameter sets. It is possible to measure
avalanches of activity by defining a small threshold
(e.g. $\theta=10^{-6}$) for the overall (integrated) activity and
analyzing the statistics of excursions (sizes S and times T)
above such a threshold \cite{Branching,SOB}. Results of our extensive
computer simulations are shown Fig.\ref{fig:ava} and
Fig.\ref{fig:xis}.

In particular, Fig.\ref{fig:ava} reveals the existence of
scale-invariant episodes of activity --whose distributions are very
well fitted by the exponents of the unbiased branching process--
appearing together with anomalous ``king'' avalanches in full analogy
with the theory of SOB \cite{SOB}. These computational results are
very robust to changes in the control parameter $\xi$ as illustrated
in Fig.\ref{fig:xis}. Actually, hardly any difference is observed in
the probability distributions when $\xi$ is increased from a value $1$
to $10$.  This reveals the emergence of true self-organization
regardless of the specific value of the parameters in the limit in
which an infinite separation of timescales is imposed.

Thus, at the steady state, the system with infinitely-separated
timescales actually self-organizes to the edge of a
discontinuous/first-order phase transition, where active and absorbing
phases coexist and scale-invariant episodes of activity emerge,
  as in SOB.

Finally, observe that the lack of a diffusion term in the equation for
synaptic resources does not seem to be a problem, indicating that
diffusion in the second equation is not essential (``relevant'' in the
jargon of the renormalization group) feature and that possibly it
could also be removed from the minimal set of Eqs.(1) describing SOB
(this claim is also supported by our own computational analyses).

\section{Discussion}

We have shown, by using the recently proposed Landau-Ginzburg model of
cortex dynamics \cite{PNAS-Synchro}, that self-organization behavior
to the edge of a discontinuous phase transition with bistability can
be recovered by considering extremely slow synaptic timescales.

First, from a theoretical point of view, our analyses reveal that
different regulatory mechanisms for the control parameter,
--i.e. different equations for the ``energy'' field, with different
meanings and possibly with diverse features such as conserved dynamics
or not, or with or without a diffusion term-- can be considered in the
context of self-organization to the neighborhood of a discontinuous
transition.  The SOB phenomenology --which is characterized by
scale-free avalanches controlled by branching process exponents,
appearing together with anomalously-large avalanches-- emerges once 
the ``self-organization limit'' is taken: in the case considered here
this means that the dynamics of charging and discharging synapses
needs to be infinitely slow.

Our approach allows us to illustrate that, one can use diverse
dynamical models to recover scale-free distributions of neuronal
avalanches --with the empirically observed branching-process
exponents-- but, the synaptic timescales needed to recover them in the
limit of SOB are not compatible with those of a realistic short term
plasticity dynamics.  Indeed, as reported by neurophysiological
measurements, realistic synaptic timescales are comprised between few
hundreds of milliseconds up to few seconds
\cite{Vives,Buzsaki,Tsodyks,Abbott1997}, which are clearly far from
the SOB limit, requiring a much slower dynamics.  For instance, the
unit of time, in the Landau-Ginzburg model, should be understood in
terms of the spontaneous decay of neural activity: one time unit in Eq
(\ref{PNAS}) is of the order of the millisecond, which means that in
order to recover the SOB limit, the recovery timescale for synaptic
resources should be of the order of several minutes!  On the other
hand, the Landau-Ginzburg model --leading to verifiable predictions--
is based on the consideration of realistic timescales.  In other
words, the ``imperfect'' form of SOB, with finite time scales is much
moire adequate to describe neural dynamics than SOB itself.

Thus, self-organization to the edge of bistability can be achieved
only by considering an infinite separation of timescales that does not
seem to be plausible in the cortex, at least not by considering
short-term plasticity as a chief regulatory mechanism.  Importantly,
the same criticism can be made --and has been explicitly made
\cite{JABO2}-- to models for neural dynamics based on self-organized
criticality: they also require an infinite separation of timescales to
work. In order to circumvent this conceptual problem in SOC, i.e. to
deal with real systems that are not perfectly conservative nor operate
under the strict condition of perfectly separated timescales --as
mentioned above-- researchers introduced the concept of self-organized
quasi criticality (SOqC). Such quasi-critical systems self-organize to
operate within a neighborhood of the critical point, with continuous
excursions to both sides. In other words, when the strict mathematical
conditions for SOC are relaxed, the resulting self-organized systems
are not perfectly critical nor show perfect scale invariance. Rather
they hover around the critical point and exhibit imperfect scaling
\cite{JABO1}.

Thus, in full analogy with SOqC, we propose that in order to describe
``imperfect'' SOB systems, such as the Landau Ginzburg theory
discussed here, one should use a terminology, such as ``self-organized
\emph{quasi} bistability'', or better, ``self-organized collective
oscillations'': if the separation of timescales is not large, then
--depending on the control parameter value-- the system is either in a
down state, up state, or in an intermediate range of large waves or
collective oscillations. However, as the separation of timescales is
progressively increased, the system becomes more and more SOB-like,
with scale-free avalanches and progressively less frequent large waves
of activity (see \cite{PNAS} as well as \cite{Burioni1,Burioni2}).

From the viewpoint of neuroscience, it is important to recall that the
original Tsodyks-Markram model for short-time synaptic plasticity
implements an additional variable describing a ``facilitation''
mechanism whose timescale could be much longer than the depression one
considered here \cite{Tsodyks,Abbott1997,Cortes2013}.  This would make
the dynamics more complex and eventually regulate the baseline of
synaptic resources, allowing self-tuning to the transition point.
Also, one could explore the role played by other types of
plasticities, including facilitation, such as spike-timing dependent
plasticity.  In particular, plasticity operates in the cortex through
a wide spectrum of timescales.  For example, long term potentiation is
known to involve different mechanisms, such as the production of new
ion channels, modifying the excitability of the units for a
long-lasting period. It is likely that for homeostatic mechanisms with
larger separation of timescales, behavior closer to SOB could emerge.

Actually, in order to illustrate the generality of the discussions,
here we have conducted similar analyses on a similar model for neural
dynamics in the presence of \emph{adaptation} (rather than synaptic
plasticity), recently introduced by Levenstein \emph{et al.}  (see
Appendix C).  This research line seems promising and we leave a
careful exploration of this possibility for a future work.

Summing up, self-organized bistability with its concomitant scale-free
avalanches of activity can be obtained as a limiting case of the
Landau-Ginzburg model for cortex dynamics. However, this only occurs
in the unrealistic limit of an extremely slow dynamics for synaptic
resources is considered. On the other hand, relaxing the strict
conditions for SOB (i.e. not imposing a huge separation of
timescales) one obtains a version of the SOB theory --that have also
very remarkable features-- and that we call self-organized quasi
bistability, or better, ``\emph{self-organized collective
  oscillations}'', that  can be much more adequate to describe real
systems, e.g. in neuroscience.

\vspace{1cm}

\begin{acknowledgments}
  We acknowledge the Spanish Ministry and Agencia Estatal de
  investigaci{\'o}n (AEI) through grant FIS2017-84256-P (European
  Regional Development Fund), as well as the Consejer{\'\i}a de
  Conocimiento, Investigaci{\'o}n y Universidad, Junta de
  Andaluc{\'\i}a and European Regional Development Fund (ERDF),
  ref. $SOMM17/6105/UGR$ for financial support.  We also thank
  Cariparma for their support through the TEACH IN PARMA project.
\end{acknowledgments}

\appendix

\section{Mean-field approach to SOB}

\begin{figure}
\centering
\includegraphics[width=\columnwidth]{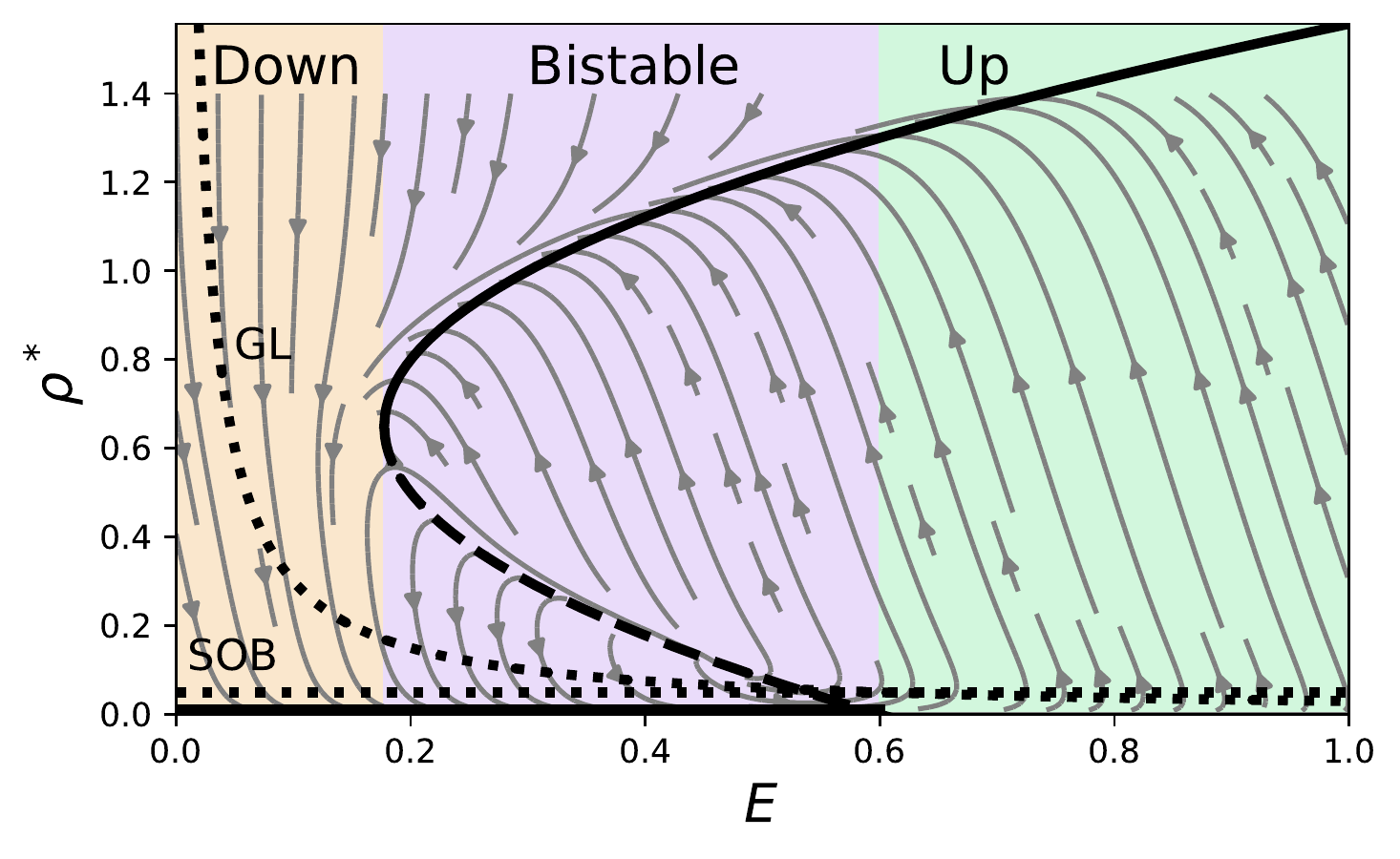}
\caption{{\bf Two forms of self-organization.} Diagram showing the
  steady-state solutions $\rho^*$ of Eq.(\ref{activity}) as a function
  of $E$. The solid line indicates stable fixed points, while the
  dashed line indicates unstable ones. Taking $E$ as a control
  parameter there are three different regions: a down state, a
  bistability region, and an up state. When a dynamics for $E$ is
  added, the system may display oscillations. For the sake of
  comparison, nullclines are displayed for both the SOB,
  Eq.(\ref{SOB-MF}) (horizontal dotted line) and the Landau-Ginzburg
  for the cortex, Eq.(\ref{xi}) (curved dotted line) models. Note that
  the nullclines in these two cases tend to cut tangentially at the
  bifurcation point, $\xi=a$, --becoming locally indistinguishable--
  as the large timescale separation limit is approached. Gray arrows
  represent the corresponding vector field for the case of the
  Landau-Ginzburg model.  Parameters have been set to
  $a =0.6, b=1.3, h=0.005, \varepsilon=0.1, \xi=1.0$. \label{fig:rho}}
\end{figure}

Here we perform a detailed single-site or mean-field analysis of the
dynamical system describing SOB, i.e. Eqs.(\ref{SOB}) but neglecting
spatial dependence and noise. The first equation in Eqs.(1) can then
be simply written as
\begin{equation}
\dot \rho = (-a+E)\rho+b\rho^2-\rho^3 +h 
\label{activitymf}
\end{equation}
where now $\rho$ is the overall density of activity.  Let us analyze
this equation for $\dot E=0$, i.e. for a fixed value of $E$, with
vanishing driving $h$ and dissipation $\varepsilon$. Then
Eq.(\ref{activitymf}) has the typical form of a discontinuous
transition for $\rho$. Actually, imposing stationarity, $\dot\rho = 0$, one obtains
three possible solutions: $\rho_0 = 0$ (stable for $E < a$) and a
positive and negative pair,
\begin{equation}
\rho_\pm = \frac{b}{2} \pm \sqrt{\frac{b^2}{4}+(-a+E)}
\label{activity}
\end{equation}
emerging at a saddle-node bifurcation. As illustrated in
Fig.\ref{fig:rho}, these last ones exist (as real solutions) only for
$E > a - b^2/4$ and $\rho_+$ is stable for $a - b^2/4 < E < a$. On the
other hand, $\rho_-$ is unstable in this region, and becomes
non-physical, i.e. negative for $E > a$. This gives a bistability
region (shaded in magenta color in Fig.\ref{fig:rho}) where two stable
fixed points $\rho_0$ and $\rho_+$ coexist, with a branch of unstable
fixed points $\rho_-$ in between.

These are the solutions of the problem for constant ``energy''
$E$. For the full system, including the dynamics for $E$
(i.e. for $h\neq0$, $\varepsilon\neq0$) the mean-field dynamics of the
control parameter $E$ is given by:
\begin{equation}
\dot E = h - \varepsilon \rho.
\label{SOB-MF}
\end{equation}
Observe that this dynamics acts differentially depending on the phase,
i.e. on the value of $\rho$: in the absence of activity it increases
owing to the driving $h$,  while as activity grows the
  dissipation (negative) dominates the dynamics.  Observe also that
the only possible fixed point --where these two contributions
balance-- is $\rho^* = h / \varepsilon$. Observe that, this equation
defines a nullcline which is a horizontal line in Fig.\ref{fig:rho}.
Note also that in the double limit of conservation
$h,\varepsilon \rightarrow 0$ with an infinite separation of
timescales $h/\varepsilon \rightarrow 0$, this is necessarily a point
of vanishingly small activity and, as seen above (see
Fig.\ref{fig:rho}), the branch of points with low activity is
unstable.  Indeed, from Eq.(\ref{activity}) at stationarity one
obtains $E^* = a - bh/\varepsilon + (h/\varepsilon)^2$, which
corresponds to an unstable spiral focus. Inspection of the associated
velocity field reveals that the dynamics exhibits a stable limit
  cycle around such an unstable fixed point (see Fig.\ref{fig:rho}),
shifting cyclically between the upper and the lower branches, with
high and low activity, respectively. Note that in the limit
$h/\varepsilon \rightarrow 0$, the timescale between succesive cycles
increases (in particular, the imaginary part of the eigenvalues
associated with the unstable spiral $(\rho^*, E^*)$, scales as 
  $ \sim (h/\varepsilon)^{-1} $, meaning that the frequency of
oscillations goes to $0$ as happens in an infinite period bifurcation
\cite{Strogatz}).

This simple mean-field approach shows that the system behaves as an
\emph{excitable system} in which following external driving there
  are cycles of periodic activity. As shown in detail in \cite{SOB},
going beyond mean-field --i.e. turning back to a spatially-explicit
system such as a lattice-- this mechanism generates scale-free
avalanches alternating with system-wide spanning waves of
activity. The reason behind such a change is that in the above
situation, driving may induce local jumps of the activity from the
lower branch to the upper one, and from this it can propagate to
nearest neighbors coupled to it, giving rise to a local avalanche of
activity \cite{SOB}.

\section{Mean-field approach to the Landau-Ginzburg theory}

Here, we discuss the single site or mean-field description of the
Landau-Ginzburg theory for cortex dynamics. The first single-unit
equation is identical to Eq.\ref{activitymf} while the second one
reads
\begin{equation}
\dot E = \frac{1}{\tau_R}(\xi-E)- \frac{1}{\tau_D} E\rho.
\label{xi}
\end{equation}
where $E$ is now the overall ``energy'' or density of synaptic
resources. We call it here $E$ rather than $R$ as in Eq.(\ref{PNAS})
to make a more direct comparison with Appendix A; also to make the
parallelism with the SOB theory even more explicit, we define
$h = 1/\tau_R$ and $\varepsilon = 1/\tau_D$, and study the double
limit of slow synaptic dynamics $h,\varepsilon \rightarrow 0$ with
$ \Delta \equiv h/\varepsilon \rightarrow 0$, in which the two
synaptic timescales are infinitely separated.

\begin{figure}
\centering
\includegraphics[width=\columnwidth]{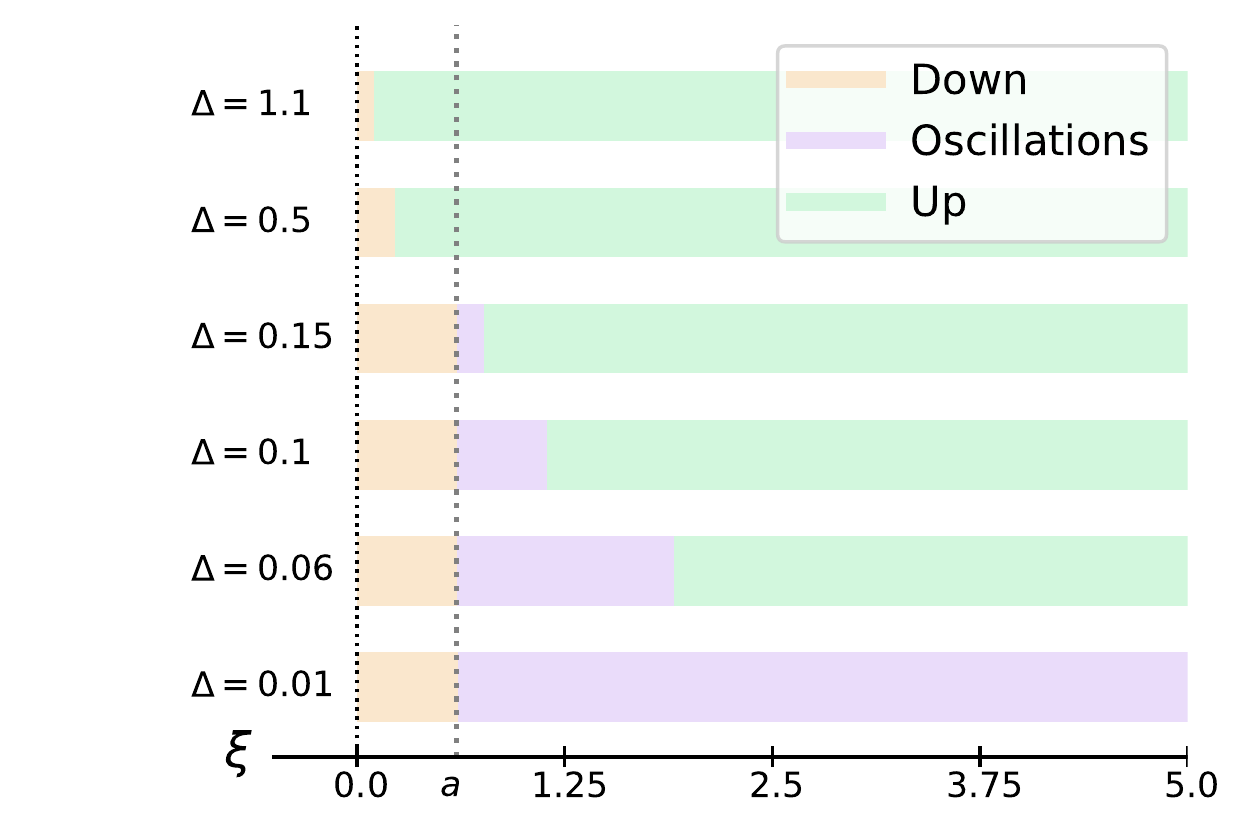}
\caption{{\bf Invariance on $\xi$.} Behaviour of the Landau-Ginzburg
  system as a function of the control parameter $\xi$, as the limit
  $\Delta \rightarrow 0$ is approached. Each color represents a
  different phase (see legend). Note that as the separation of
  timescales increases (i.e. as $\Delta \rightarrow 0$, in the SOB
  limit) the oscillatory phase takes more space until it comprises the
  whole region $\xi > a$. Parameter values: $a =0.6, b=1.3$.}
\label{inv}
\end{figure}

The main difference between this mean-field theory and its SOB
counterpart (in Appendix A) is the existence in Eq.(\ref{xi}) of a key
parameter $\xi$, which bounds the largest possible value of $E$.
Three cases can be distinguished depending on the value of $\xi$ and
the separation of timescales.

(i) If $\xi < a$ obviously $E < a$, and then equation (\ref{activity})
has a negative linear term, implying that $\rho = 0$ is the only
possible stable state. In other words the two nullclines intersect in
the absorbing phase, i.e. for $\rho=0$ (see orange line in Fig.\ref{fig:mf}a).

(ii) More in general, imposing $\dot E = 0$, one finds the fixed point
value $E^* = \frac{h\xi}{(h+\varepsilon)\rho}$ (defined only for
$\rho > 0$) which in the limit of large timescale separation becomes
$E^* \simeq \frac{\Delta \xi}{ \rho}$, implying that the nullclines
intersect at an unstable fixed point (see red line in the leftmost
inset of Fig.\ref{fig:mf}), after a homoclinic bifurcation at $E=a$. This leads
to oscillations in the following way: within the absorbing state,
energy slowly increases as $\dot{E}=h(\xi - E)$ and --if $\xi > a$--
at some point the system reaches the value $E=a$ where the absorbing
state $\rho=0$ becomes unstable and activity grows.  Then, the energy
fixed point is given by $E^*$. If the energy is larger than this
value, it decreases in a fast way, until $E=E^*$. However, if $\Delta$
is very small, the fixed point is $E^* \simeq 0$, meaning that $E$
decreases until the up-state branch becomes unstable again -falling to
the absorbing state. When the absorbing state is reached, $E^*$ is
unstable and the cycle starts again. This simple back-and-forth
mechanism is responsible for the emergence of limit cycles in the
dynamics for $\xi > a$ (see also \cite{Sisyphus} where a mechanism
very similar to this is termed ``Sisyphus effect''). This type of
effect is well-known in the theoretical-neuroscience literature (see
e.g. \cite{Vives,latham1,Bazhenov2002,Compte2003,Giugliano2004,Gigante,Curto2009,Tabak2011,Levenstein2019}).

(iii) For larger values of $\xi$, such that the nullclines intersect
in the (stable) up-state branch (see the brown line in the leftmost
inset of Fig.\ref{fig:mf}) there are no oscillations and the system sets in an
active state with $E^*= \frac{\Delta \xi}{ \rho}$ (the oscillations
disappear at a Hopf bifurcation).  Observe that the condition for this
case (iii) to emerge depends on the separation of timescales,
$\Delta$. In particular, in the limit $\Delta \rightarrow 0$, a value
$\xi \rightarrow +\infty$ is required to destabilize the cycle limit.
This effect is illustrated in Figure \ref{inv}, where it can be seen
that as the $\Delta \rightarrow 0$ limit is approached the regime of
oscillations broadens (up to infinite).

Finally, let us note that in case (ii) --which is the one for which
avalanches are obtained once the spatially explicit version of the
model is considered-- the nullcline associated with Eq.(\ref{xi})
becomes very similar to its counterpart for the SOB case: both of
  them intersect the unstable branch of $\rho$ solutions in the same
  part of it and with flat slopes in both cases in the limit
  $\Delta \rightarrow 0$.  Also, observe in the upper inset of Fig.\ref{fig:mf}
that the nullclines are quite insensitive to the specific value of
$\xi$ (as long as $\xi >a$) when such a limit is taken.  Thus, the
same phenomenology is expected to emerge in both cases.

\begin{figure}[h]
\centering
\includegraphics[width=\columnwidth]{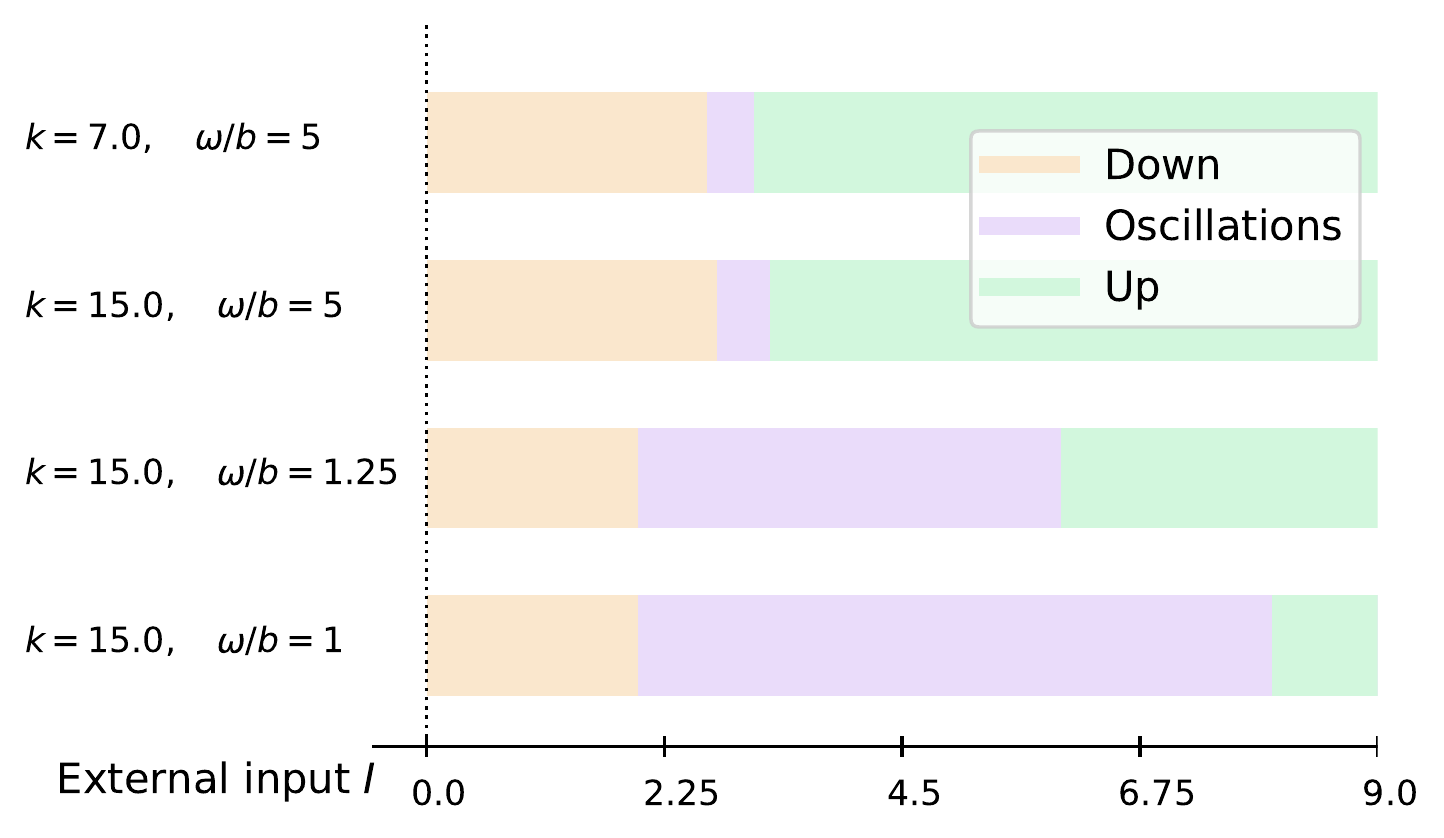}
\caption{{\bf Width of the region in which collective oscillations
    emerge in the model for adaptation model of Levenstein \emph{et
      al.}\cite{Levenstein2019}.} It shows that either increasing $k$
  or, alternatively, decreasing the ratio $\omega/b $ enhances the
  region of collective oscillation, as in Figure \ref{inv}. Taking,
  either of these combinations of parameters to their limit
  $k\rightarrow \infty$ or $\omega/b \rightarrow 0$ one observes a
  huge range for self-organized collective oscillations.}
\label{inv-levenstein}
\end{figure}

\section{Neural model with adaptation}

The ideas exposed above can be applied to models of neural activity
relying on regulatory mechanisms other than synaptic plasticity. To
illustrate this, here we consider a model recently proposed by
Levenstein \emph{et al.} \cite{Levenstein2019} to describe cortical
dynamics. The model is defined by two differential equations, one for
the spike rate $r$ of a cortical mesoscopic region and another one for
an ``adaptation variable'' $a$ that regulates the spiking rate; in
particular:
\begin{eqnarray}
\dot r &=& -r + S ( \omega r -b a + I, 1, I_0), \\ 
\dot a &=& -a + S (r, k, r_0), \label{Levenstein}
\end{eqnarray}
where $S(x,k,x_0) = 1 / \left( 1 + \exp(-k(x-x_0)) \right)$ is a
sigmoidal function whose steepness is controlled by the parameter
$k$, $x_0$ is a threshold, $I$ the external input, and $\omega$ and
$b$ are the coupling and the adaptation strengths, respectively.

The equation for the mean rate is very similar to the corresponding
one for the Landau-Ginzburg theory Eq.(\ref{activitymf}); actually,
truncating the series expansion of $S$ around the threshold $I_0$,
transforms the first equation into a third-degree polynomial,
exhibiting also a discontinuous transition as Eq.(\ref{activitymf}).

To ease the comparison between Eq.(\ref{Levenstein}) and the
Landau-Ginzburg theory given by Eq.(\ref{PNAS}), we perform a change
of variables $E=1/a$. Expanding in power series in $r-r_0$ and
truncating up to leading order, the adaptation equation becomes
$\dot E = E \left( 1 - E (2+k(r-r_0))/4 \right)$.  Observe that here
$E=1/a$ plays a role analogous to that of the synaptic resources in
the Landau-Ginzburg theory, and that the similitude between both
theories (at a mean-field level) is self-evident.

Notice also that the energy charge (positive term) is of order unity,
while the discharge (negative) grows with $k$; thus, taking the limit
$k \gg 1$ of very steep adaptation  and taking the whole adaptive
dynamics to be arbitrarily slow, leads to similar effect as the
infinite separation of timescales limit in the case of the
Landau-Ginzburg theory.

An alternative,  more efficient, way to achieve the limit of self-organization is to
modify the coupling and the adaptation strength variables ($\omega$
and $b$). Actually, our Figure \ref{fig:mf} is very similar to
Fig.3B in \cite{Levenstein2019}, where the role of the timescales
$\tau_R$ and $\tau_D$ is played by $\omega$ and $b$. This implies that
taking the limit $\omega / b \rightarrow 0$ while increasing both
variables should enlarge the oscillatory region in the same way that
happens with the non-conserved SOB and Landau-Ginzburg theory (see
Appendices A and B). However, the SOB-limit of perfect  self-organization
cannot be reached in this model, as there is no absorbing or quiescent
state.

We performed numerical simulations of Eq.(\ref{Levenstein}) and
measured the length of the oscillatory region as a function of
parameters ($k$ on the one hand or $\omega$, and $b$ on the other). We
find --as illustrated in Figure \ref{inv-levenstein}-- that the width of
such a region increases as the separation of timescales is enlarged
(much as in Fig.\ref{inv}).

Thus, self-organized collective oscillations emerge in a rather
generic way for reasonable parameter choices in this model. This
result illustrates that the ideas discussed before for neural dynamics
and self-organization are not specific of the Landau-Ginzburg theory
relying on short-term synaptic plasticity, but can be also extended
to other models of neural dynamics.

%

\end{document}